\documentstyle[12pt]{article}

\begin{document}

\title{\Large {\bf Microcausality violation of scalar field on
       noncommutative spacetime with the time-space noncommutativity } }

\author{{{Zheng Ze Ma}} \thanks{E-mail address:
           z.z.ma@seu.edu.cn}
  \\  \\ {\normalsize {\sl Department of Physics, Southeast University,
           Nanjing, 210096, P. R. China } } }

\date{}

\maketitle

\vskip 0.8cm

\centerline{\normalsize {\bf Abstract ~~~ }}

\vskip 0.5cm

\begin{minipage}{14.5cm}

\indent

\baselineskip 16 pt

~~~~  Quantum field theories on noncommutative spacetime have many
different properties from those on commutative spacetime. In this
paper, we study the microcausality of free scalar field on
noncommutative spacetime. We expand the scalar field in the form of
usual Lorentz invariant measure in noncommutative spacetime. Then we
calculate the expectation values of the Moyal commutators for the
quadratic operators, such as $\varphi(x)\star\varphi(x)$, $\pi({\bf
x},t)\star\pi({\bf x},t)$, $\partial_{i}\varphi({\bf
x},t)\star\partial_{i}\varphi({\bf x},t)$, and
$\partial_{i}\varphi({\bf x},t)\star\pi({\bf x},t)$. We obtain that
for space-space noncommutativity, i.e., $\theta^{0i}=0$ in
$\theta^{\mu\nu}$, microcausality of free scalar field on
noncommutative spacetime is satisfied. For time-space
noncommutativity, i.e., $\theta^{0i}\neq0$ in $\theta^{\mu\nu}$,
microcausality of free scalar field on noncommutative spacetime is
violated.

\end{minipage}

\vskip 0.5cm

PACS numbers: 11.10.Nx, 03.70.+k

\vskip 0.5cm

\baselineskip 16 pt

\section{Introduction}

\indent

  Many years ago, the model of quantized and noncommutative spacetime
was first constructed by Snyder [1]. The mathematical development on
noncommutative geometry was carried out by Connes [2]. In Ref. [3],
Doplicher {\sl et al}. proposed the uncertainty relations for the
measurement of spacetime coordinates from the Heisenberg's
uncertainty principle and Einstein's gravitational equations. In
recent years, spacetime noncommutativity was discovered again in
superstring theories [4]. It has resulted a lot of researches on
noncommutative field theories (NCFTs) [5,6].

  The coordinates of noncommutative spacetime become noncommutative
operators. They satisfy the commutation relations
$$
  [{\bf x}^{\mu},{\bf x}^{\nu}]=i\theta^{\mu\nu} ~,
  \eqno{(1.1)}  $$
where $\theta^{\mu\nu}$ is a constant real antisymmetric matrix with
the dimension of square of length. Field theories on noncommutative
spacetime can be formulated through the Weyl-Moyal correspondence
[5,6]. Every field $\phi(\bf x)$ defined on noncommutative spacetime
is mapped to its Weyl symbol $\phi(x)$ that defined on the
corresponding commutative spacetime. At the same time, the products
of field functions are replaced by the Moyal star-products of their
Weyl symbols
$$
  \phi({\bf x})\psi({\bf x})\rightarrow \phi(x)\star\psi(x) ~,
  \eqno{(1.2)}  $$
where the Moyal star-product is defined by
\begin{eqnarray*}
~
  \phi(x)\star\psi(x) & = & e^{\frac{i}{2}\theta^{\mu\nu}\frac{\partial}
                  {\partial
                 \alpha^\mu}\frac{\partial}{\partial\beta^{\nu}}}
                 \phi(x+\alpha)
                 \psi(x+\beta)\vert_{\alpha=\beta=0} \\
                 & = & \phi(x)\psi(x)+\sum\limits^{\infty}_{n=1}
                 \left(\frac{i}{2}\right)^{n}
                 \frac{1}{n!}\theta^{\mu_{1}\nu_{1}}
                 \cdots\theta^{\mu_{n}\nu_{n}}
                 \partial_{\mu_{1}}\cdots\partial_{\mu_{n}}\phi(x)
                 \partial_{\nu_{1}}\cdots\partial_{\nu_{n}}\psi(x) ~.
~~(1.3)
\end{eqnarray*}
Thus, the Lagrangians of field theories on noncommutative spacetime
can be formulated directly through replacing the products by the
Moyal star-products in the Lagrangians of field theories on
commutative spacetime. And the commutators of coordinate operators
of Eq. (1.1) are equivalently replaced by the Moyal star-product
commutators of the noncommutative coordinates
$$
  [x^{\mu},x^{\nu}]_{\star}=i\theta^{\mu\nu} ~.
  \eqno{(1.4)}  $$

  Quantum field theories on noncommutative spacetime have many
different properties from those on commutative spacetime [5,6]. In
this paper, we study the microcausality of scalar field on
noncommutative spacetime. In fact, such a problem need to be studied
from several different aspects.

  First, such a problem is related with the Lorentz invariance
problem of NCFTs. It is well known that NCFTs violate Lorentz
invariance because of their theoretical constructions. As pointed
out in Ref. [7], there are two kinds of Lorentz transformations for
NCFTs. One is the observer Lorentz transformation. In this case, one
can suppose $\theta^{\mu\nu}$ carries Lorentz indices, which means
that $\theta^{\mu\nu}$ transforms covariantly under the
transformations of observer's frame. This will leave the physics
unchanged because both field operators and $\theta^{\mu\nu}$
transform covariantly. The other is the particle Lorentz
transformation. In this case, Lorentz transformations for field
operators are taken in a fixed observer frame. Because
$\theta^{\mu\nu}$ are not fields, they are only certain parameters
in the theory, particle Lorentz transformation leave
$\theta^{\mu\nu}$ unchanged and hence modify the physics.

  Therefore NCFTs at least violate particle Lorentz transformation
invariance. In Ref. [8], the authors have proposed that NCFTs
satisfy the twisted Poincar{\' e} invariance. In Ref. [9], the
authors have pointed out that for Lorentz transformations that leave
$\theta^{\mu\nu}$ unchanged, such as the particle Lorentz
transformation mentioned above, the $SO(3,1)$ Lorentz group is
broken down to a subgroup $SO(1,1)\times SO(2)$. Therefore in Ref.
[9], the authors constructed the $SO(1,1)\times SO(2)$ invariant
spectral measure for the expansions of quantum fields on
noncommutative spacetime. They have proposed that microcausality
condition for quantum fields on noncommutative spacetime should be
formulated with respect to the two dimensional light-wedge.
Therefore the traditional concept of microcausality is violated
generally for quantum fields on noncommutative spacetime [9-12].

  On the other hand, corresponding to the observer Lorentz
transformations for NCFTs, one should expand quantum fields on
noncommutative spacetime in the form of their usual Lorentz
invariant measures in order to study their microcausality
properties. In fact, for such a problem, some results were given in
Ref. [13]. Recent results were obtained by Greenberg [14]. In Ref.
[14], Greenberg have obtained that for scalar field on
noncommutative spacetime,
$[:\varphi(x)\star\varphi(x):,:\varphi(y)\star\varphi(y):]_{\star}$
is nonzero for a spacelike interval for the case $\theta^{0i}\neq 0$
and $\theta^{0i}=0$. Thus microcausality is violated for scalar
field on noncommutative spacetime generally. In this paper we will
study this problem further. We obtain the results different from
those of Ref. [14].

  Besides, we need to investigate microcausality properties of
interacting fields on noncommutative spacetime. Recently in Ref.
[15], Haque and Joglekar have obtained that for the Yukawa
interaction in noncommutative spacetime, causality is violated for
both $\theta^{0i}\neq 0$ and $\theta^{0i}=0$. In addition we can see
from Refs. [16-19] that quantum and classical nonlinear
perturbations have infinite propagation speed in noncommutative
spacetime. While these phenomena should also have relations with the
violation of causality of quantum fields on noncommutative
spacetime.

  Because in the following we need to calculate the commutation
relations of two field operators defined at two different spacetime
points, we need to generalize Eq. (1.3) for two fields defined at
two different spacetime points. In fact, such a formula has been
given in Ref. [6]. It reads
\begin{eqnarray*}
  \phi(x_{1})\star\psi(x_{2}) & = & e^{\frac{i}{2}
          \theta^{\mu\nu}\frac{\partial}{\partial
          \alpha^\mu}\frac{\partial}{\partial\beta^{\nu}}}\phi(x_{1}+\alpha)
                 \psi(x_{2}+\beta)\vert_{\alpha=\beta=0} \\
                 & = & \phi(x_{1})\psi(x_{2})+\sum\limits^{\infty}_{n=1}
                 \left(\frac{i}{2}\right)^{n}
                 \frac{1}{n!}\theta^{\mu_{1}\nu_{1}}
                 \cdots\theta^{\mu_{n}\nu_{n}}
                 \partial_{\mu_{1}}\cdots\partial_{\mu_{n}}\phi(x_{1})
                 \partial_{\nu_{1}}\cdots\partial_{\nu_{n}}\psi(x_{2}) ~.
\end{eqnarray*}
$$  \eqno{(1.5)}  $$
This formula can be derived through the standard {\sl Weyl symbol}
method as demonstrated in Ref. [6], supposing that we generalize the
spacetime commutation relations (1.1) to two different spacetime
points
$$
  [{\bf x}_{1}^{\mu},{\bf x}_{2}^{\nu}]=i\theta^{\mu\nu} ~.
  \eqno{(1.6)}  $$
A simplified derivation for Eqs. (1.5) and (1.6) has been given by
Chaichian {\sl et al}. in Ref. [20] using the concept of quantum
shift. Consequently, the Moyal star-product commutators for the
noncommutative coordinates of Eq. (1.4) become
$$
  [x_{1}^{\mu},x_{2}^{\nu}]_{\star}=i\theta^{\mu\nu} ~,
  \eqno{(1.7)}  $$
which can be resulted from Eq. (1.5). Equation (1.5) has been used
many where in the literature [14,15,21].

  The content of this paper is organized as follows. In Sec. 2, we
analyze the measurement of quantum fields on noncommutative
spacetime and the criterion of microcausality violation. In Sec. 3,
we calculate the vacuum state expectation value for the Moyal
commutator
$[\varphi(x)\star\varphi(x),\varphi(y)\star\varphi(y)]_{\star}$ and
obtain that microcausality is violated for the case
$\theta^{0i}\neq0$ of spacetime noncommutativity. In Sec. 4, we
calculate the non-vacuum state expectation value for the Moyal
commutator
$[\varphi(x)\star\varphi(x),\varphi(y)\star\varphi(y)]_{\star}$ and
obtain that microcausality is violated for the case
$\theta^{0i}\neq0$ of spacetime noncommutativity. In Sec. 5, we
calculate the expectation values for some other quadratic operators
of scalar field on noncommutative spacetime and obtain the similar
result as that of Sec. 3 and Sec. 4. In Sec. 6, we discuss some of
the problems.

\section{The criterion of microcausality violation}

\indent

  In this section, we first analyze the measurement of quantum fields
on noncommutative spacetime and the criterion of microcausality
violation.

  For quantum field theories, as well as quantum mechanics, what the
observer measures are certain expectation values. We suppose that there
are two observers A and B situated at spacetime points $x$ and $y$, they
proceed a measurement separately on the state vector $\vert\Psi\rangle$
for the locally observable quantity ${\cal O}(x)$ in the same occasion.
However the time $x_{0}$ may not equal to the time $y_{0}$ generally.
For the observer A, the state vector $\vert\Psi\rangle$ has been affected
by the measurement of the observer B at the spacetime point $y$. Or we
can say the observer B's observation instrument has taken an action on
the state vector $\vert\Psi\rangle$. The state vector has become
${\cal O}(y)\vert\Psi\rangle$. When the observer A takes his or her
measurement on the state vector, his or her observation instrument
will act on the state vector ${\cal O}(y)\vert\Psi\rangle$ again.
These two sequent actions should be represented by the product
operation of the operators. However because now the spacetime is
noncommutative, the product operation should be the Moyal star-product,
while not the ordinary product. Or we regard that in noncommutative
spacetime, the basic product operation is the Moyal star-product.
Thus what the measuring result the observer A obtained from his or her
instrument is
$\langle\Psi\vert{\cal O}(x)\star{\cal O}(y)\vert\Psi\rangle$.
Similarly for the observer B, the state vector $\vert\Psi\rangle$
has been affected by the action of the observer A's instrument at
the spacetime point $x$. The state vector becomes
${\cal O}(x)\vert\Psi\rangle$. What the measuring result the
observer B obtained from his or her instrument is
$\langle\Psi\vert{\cal O}(y)\star{\cal O}(x)\vert\Psi\rangle$.

  Supposing that microcausality is satisfied for NCFTs, this means
that there do not exist the physical information and interaction
with the transmission speed faster than the speed of light. Thus
when the spacetime interval between $x$ and $y$ is spacelike, the
affection of the observer B's measurement or the action of observer
B's instrument at spacetime point $y$ on the state vector
$\vert\Psi\rangle$ has not propagated to the spacetime point $x$
when the observer A takes his or her measurement on the state vector
$\vert\Psi\rangle$ at the spacetime point $x$. These two physical
measurements do not interfere with each other. For the observer A,
the state vector is still $\vert\Psi\rangle$, while not ${\cal
O}(y)\vert\Psi\rangle$. Therefore the measuring result what the
observer A obtained is just $\langle\Psi\vert{\cal
O}(x)\vert\Psi\rangle$. Thus from the sense of experimental
measuring discussed above, we have
$$
  \langle\Psi\vert{\cal O}(x)\star{\cal O}(y)]\vert\Psi\rangle=
     \langle\Psi\vert{\cal O}(x)\vert\Psi\rangle
  ~~~~~~ \mbox{for} ~~~~~~ (x-y)^{2}<0 ~.
  \eqno{(2.1)}  $$
The same reason as the observer A, the measuring result what the
observer B obtained at the spacetime point $y$ is just
$\langle\Psi\vert{\cal O}(y)\vert\Psi\rangle$. Thus from the sense
of experimental measuring discussed above, we have
$$
  \langle\Psi\vert{\cal O}(y)\star{\cal O}(x)]\vert\Psi\rangle=
     \langle\Psi\vert{\cal O}(y)\vert\Psi\rangle
  ~~~~~~ \mbox{for} ~~~~~~ (x-y)^{2}<0 ~.
  \eqno{(2.2)}  $$
We can suppose that the state vector $\vert\Psi\rangle$ is in the
momentum eigenstate, thus it is also in the energy eigenstate.
At the same time, we can suppose that the state vector
$\vert\Psi\rangle$ is in the Heisenberg picture, therefore it does
not rely on the spacetime coordinates. From the Heisenberg
relations and the translation transformation, we have
\footnote{
  We note $y-x=a$. From the Heisenberg relations and the translation
  transformation, we have
  $$  {\cal O}(y)=\exp(ia_{\mu}{\bf P}^{\mu}){\cal O}(x)
\exp(-ia_{\mu}{\bf P}^{\mu}) ~.                     $$ Because
$a_{\mu}$ now is a constant four-vector, from Eq. (1.3) we can also
write the above expression as
$$  {\cal O}(y)=\exp(ia_{\mu}{\bf P}^{\mu})\star{\cal O}(x)
\star\exp(-ia_{\mu}{\bf P}^{\mu}) ~.               $$
  We use $P^{\mu}$ to represent the eigenvalues of the energy-momentum
  of the state vector $\vert\Psi\rangle$. Therefore we obtain
  \begin{eqnarray*}
& ~ & \langle\Psi\vert{\cal O}(y)\vert\Psi\rangle=
\langle\Psi\vert\exp(ia_{\mu}{\bf P}^{\mu})\star{\cal O}(x)
\star\exp(-ia_{\mu}{\bf P}^{\mu})\vert\Psi\rangle=
\langle\Psi\vert\exp(ia_{\mu}P^{\mu})\star{\cal O}(x)
\star\exp(-ia_{\mu}P^{\mu})\vert\Psi\rangle     \\
    & = &
    \exp(ia_{\mu}P^{\mu})\star
    \langle\Psi\vert{\cal O}(x)\vert\Psi\rangle
\star\exp(-ia_{\mu}P^{\mu}) =
\exp(ia_{\mu}P^{\mu})
\langle\Psi\vert{\cal O}(x)\vert\Psi\rangle
\exp(-ia_{\mu}P^{\mu}) =
    \langle\Psi\vert{\cal O}(x)\vert\Psi\rangle ~.
   \end{eqnarray*}
Therefore Eq. (2.3) is satisfied.}
$$
  \langle\Psi\vert{\cal O}(x)\vert\Psi\rangle=
       \langle\Psi\vert{\cal O}(y)\vert\Psi\rangle ~.
  \eqno{(2.3)}  $$
Therefore the condition
$$
  \langle\Psi\vert[{\cal O}(x),{\cal O}(y)]_{\star}\vert\Psi\rangle
    =\langle\Psi\vert{\cal O}(x)\star{\cal O}(y)\vert\Psi\rangle-
      \langle\Psi\vert{\cal O}(y)\star{\cal O}(x)\vert\Psi\rangle=0
  ~~~~ \mbox{for} ~~~~ (x-y)^{2}<0
  \eqno{(2.4)}  $$
should be satisfied for a NCFT to satisfy the microcausality.

  If microcausality is violated for a NCFT,
then there may exist the physical information and interaction with
the transmission speed faster than the speed of light. For the two
measurements of the observer A and observer B located at $x$ and $y$
separated by a spacelike interval, the affection of the observer B's
measurement at spacetime point $y$ on the state vector
$\vert\Psi\rangle$ will propagate to the spacetime point $x$ when
the observer A takes his or her measurement on the state vector
$\vert\Psi\rangle$ at the spacetime point $x$, and the affection of
the observer A's measurement at spacetime point $x$ on the state
vector $\vert\Psi\rangle$ will propagate to the spacetime point $y$
when the observer B takes his or her measurement on the state vector
$\vert\Psi\rangle$ at the spacetime point $y$. These two physical
measurements will interfere with each other. For such case, Eqs.
(2.1) and (2.2) cannot be satisfied, while we still have
$\langle\Psi\vert{\cal O}(x)\vert\Psi\rangle= \langle\Psi\vert{\cal
O}(y)\vert\Psi\rangle$ as that of Eq. (2.3). Therefore generally we
have
$$
  \langle\Psi\vert[{\cal O}(x),{\cal O}(y)]_{\star}
         \vert\Psi\rangle\neq0
  ~~~~~~ \mbox{for} ~~~~~~ (x-y)^{2}<0
  \eqno{(2.5)}  $$
for a NCFT to violate the microcausality. Thus we can judge whether
the microcausality is maintained or violated for a NCFT according to
Eq. (2.5).

  Now we suppose that ${\cal O}_{1}(x)$ and ${\cal O}_{2}(y)$ are two
different observable field operators, $x$ and $y$ are separated by a
spacelike interval, two observers A and B situate at $x$ and $y$,
and microcausality is satisfied for the field theory on
noncommutative spacetime. Supposing that the observers A and B
proceed a measurement separately on the state vector
$\vert\Psi\rangle$ for the locally observable quantities ${\cal
O}_{1}$ and ${\cal O}_{2}$ at $x$ and $y$ respectively, then from
the above analysis, from the sense of experimental measuring, we
have for the observer A
$$
  \langle\Psi\vert{\cal O}_{1}(x)\star{\cal O}_{2}(y)]
\vert\Psi\rangle=\langle\Psi\vert{\cal O}_{1}(x)\vert\Psi\rangle
  ~~~~~~ \mbox{for} ~~~~~~ (x-y)^{2}<0 ~.
  \eqno{(2.6)}  $$
And we have for the observer B
$$
  \langle\Psi\vert{\cal O}_{2}(y)\star{\cal O}_{1}(x)]
\vert\Psi\rangle=\langle\Psi\vert{\cal O}_{2}(y)\vert\Psi\rangle
  ~~~~~~ \mbox{for} ~~~~~~ (x-y)^{2}<0 ~.
  \eqno{(2.7)}  $$
Because now ${\cal O}_{1}(x)$ and ${\cal O}_{2}(y)$ are two
different operators representing two different physical observable
quantities, we have generally
$$
  \langle\Psi\vert{\cal O}_{1}(x)\vert\Psi\rangle\neq
       \langle\Psi\vert{\cal O}_{2}(y)\vert\Psi\rangle ~.
  \eqno{(2.8)}  $$
Therefore from Eqs. (2.6)-(2.8) we have
$$
  \langle\Psi\vert[{\cal O}_{1}(x),{\cal O}_{2}(y)]_{\star}\vert\Psi
\rangle\neq0 ~~~~~~ \mbox{for} ~~~~~~ (x-y)^{2}<0
  \eqno{(2.9)}  $$
generally, even if $x$ and $y$ are separated by a spacelike
interval, and the field theory satisfies the microcausality.
Therefore we cannot deduce that a NCFT violates microcausality from
Eq. (2.9) from the expectation values of the Moyal commutator of two
different operators. Thus, in order to judge whether a NCFT violates
microcausality, we need to analyze the expectation values of the
Moyal commutator of the same operator as that of Eq. (2.5).

  For quantum field theories on ordinary commutative spacetime, the
analysis for the criterion of microcausality violation is similar to
the above, except that we replace the Moyal star-products by the
ordinary products in the above equations. Usually for quantum field
theories on ordinary commutative spacetime, equation (2.4) can be
simplified to
$$
  [{\cal O}(x),{\cal O}(y)]=0
  ~~~~~~ \mbox{for} ~~~~~~ (x-y)^{2}<0
  \eqno{(2.10)}  $$
for the satisfying of the microcausality. For example for scalar
field, the fundamental commutator is
$$
  [\varphi(x),\varphi(y)]=i\Delta(x-y) ~.
  \eqno{(2.11)}  $$
It is a $c$-number function. Therefore we have
$$
  \langle\Psi\vert[\varphi(x),\varphi(y)]\vert\Psi\rangle=
    [\varphi(x),\varphi(y)] ~.
  \eqno{(2.12)}  $$
Because the commutator $[\varphi(x),\varphi(y)]$ is zero for a
spacelike interval,
$\langle\Psi\vert[\varphi(x),\varphi(y)]\vert\Psi\rangle$ is also
zero for a spacelike interval. Similarly for the quadratic operator
$\varphi(x)\varphi(x)$, we have
\begin{eqnarray*}
~~~~~~~~~~
  [\varphi(x)\varphi(x),\varphi(y)\varphi(y)] & = &
        \varphi(x)[\varphi(x),\varphi(y)]\varphi(y)+
        \varphi(y)[\varphi(x),\varphi(y)]\varphi(x)       \\
      & ~ & + [\varphi(x),\varphi(y)]\varphi(x)\varphi(y)+
               \varphi(y)\varphi(x)[\varphi(x),\varphi(y)] ~.
~~~~~~(2.13)
\end{eqnarray*}
Although the result of Eq. (2.13) is not a $c$-number function, from
the fundamental commutator of Eq. (2.11) we know that
$[\varphi(x)\varphi(x),\varphi(y)\varphi(y)]$ is zero for a
spacelike interval. Therefore
$\langle\Psi\vert[\varphi(x)\varphi(x),
\varphi(y)\varphi(y)]\vert\Psi\rangle$ is also zero for a spacelike
interval. Hence for the quadratic operator $\varphi(x)\varphi(x)$ of
the scalar field theory on ordinary commutative spacetime,
microcausality is satisfied.

  For quantum field theories on noncommutative spacetime, because of
the noncommutativity of spacetime coordinates, the Moyal commutators
are not $c$-number functions generally, as can be seen in Ref. [22]
for noncommutative scalar field and Dirac field. Thus we cannot move
away the state vectors in Eqs. (2.4) and (2.5) for the criterion of
the microcausality violation for NCFTs generally. We need to
evaluate their expectation values.

\section{Vacuum state expectation values}

\indent

  For scalar field on noncommutative spacetime, the Lagrangian for
free field is given by
$$
  {\cal L}=\frac{1}{2}\partial^{\mu}\varphi\star\partial_{\mu}\varphi-
           \frac{1}{2}m^{2}\varphi\star\varphi ~.
  \eqno{(3.1)}  $$
Its Hamiltonian density and momentum density are given by [23,24]
$$
  {\cal H}(\pi,\varphi)=
       \frac{1}{2}\Big[\pi({\bf x},t)\star\pi({\bf x},t)+
       \partial_{i}\varphi({\bf x},t)\star\partial_{i}
       \varphi({\bf x},t)+m^{2}\varphi({\bf x},t)
       \star\varphi({\bf x},t)\Big]
  \eqno{(3.2)}  $$
and
$$
  {\cal P}^{i}(\pi,\varphi)=-\frac{1}{2}\left[\pi({\bf x},t)\star
         \partial_{i}\varphi({\bf x},t)+\partial_{i}
         \varphi({\bf x},t)\star\pi({\bf x},t)\right] ~,
  \eqno{(3.3)}  $$
where $\pi({\bf x},t)=\stackrel{\cdot}{\varphi}({\bf x},t)$. In
accordance with the observer Lorentz transformation invariance of
NCFTs, which means that $\theta^{\mu\nu}$ is a tensor constant, it
transforms covariantly under the transformations of observer's
frame, we can expand the free scalar field on noncommutative
spacetime according to its usual Lorentz invariant spectral measure
[25,26]. Thus we have
$$
  \varphi({\bf x},t)=\int\frac{d^{3}k}{\sqrt{(2\pi)^{3}2\omega_{k}}}
       [a(k)e^{i{\bf k}\cdot{\bf x}-i\omega t}+
        a^{\dagger}(k)e^{-i{\bf k}\cdot{\bf x}+i\omega t}]
  \eqno{(3.4)}  $$
for the Fourier expansion of the scalar field.

  In Eq. (3.4), we take the spacetime coordinates to be noncommutative.
They satisfy the Moyal star-product commutation relations (1.4) and
(1.7). The commutation relations for the creation and annihilation
operators are given by
$$
  [a(k),a^{\dagger}(k^{\prime})]=\delta^{3}
                   ({\bf k}-{\bf k}^{\prime}) ~,                 $$
$$
  [a(k),a(k)]=0 ~,      $$
$$
  [a^{\dagger}(k),a^{\dagger}(k)]=0 ~.
  \eqno{(3.5)}  $$
They are the same as those of the commutative spacetime case. The
commutator of the Moyal star-product for the scalar field is defined
to be
$$
  [\varphi(x),\varphi(y)]_{\star}=\varphi(x)\star\varphi(y)-
              \varphi(y)\star\varphi(x) ~.
  \eqno{(3.6)}  $$
In Ref. [22], the vacuum state and non-vacuum state expectation
values for the Moyal commutator (3.6) were calculated. According to
the result of Ref. [22], microcausality is satisfied for the linear
operator $\varphi(x)$ of the free scalar field on noncommutative
spacetime. However, we need to study whether microcausality is
satisfied or not for quadratic operators of free scalar field on
noncommutative spacetime. In this section, we study the quadratic
operator $\varphi(x)\star\varphi(x)$, to see whether it satisfies
the microcausality or not. Because we consider that in
noncommutative spacetime, the basic product operation is the Moyal
star-product, we need to study the commutators of Moyal
star-products.

  The Moyal commutator of the field operators
$\varphi(x)\star\varphi(x)$ and $\varphi(y)\star\varphi(y)$
is given by
\begin{eqnarray*}
~~~~~~~~~~~~~~~~~~
  & ~ & [\varphi(x)\star\varphi(x),\varphi(y)
                          \star\varphi(y)]_{\star}         \\
  & = & \varphi(x)\star[\varphi(x),\varphi(y)]_{\star}
        \star\varphi(y)+\varphi(y)\star
        [\varphi(x),\varphi(y)]_{\star}\star\varphi(x)     \\
  & ~ & +[\varphi(x),\varphi(y)]_{\star}\star\varphi(x)
         \star\varphi(y)+\varphi(y)\star\varphi(x)
         \star[\varphi(x),\varphi(y)]_{\star}              \\
  & = & \varphi(x)\star\varphi(x)\star\varphi(y)\star\varphi(y)-
        \varphi(y)\star\varphi(y)\star\varphi(x)\star\varphi(x) ~.
~~~~~~~~~~~~~~~~~~(3.7)
\end{eqnarray*}
Because the fundamental Moyal commutator
$[\varphi(x),\varphi(y)]_{\star}$ is not a $c$-number function, as
shown in Ref. [22], we need to calculate the expectation value for
Eq. (3.7) in order to investigate whether
$\varphi(x)\star\varphi(x)$ satisfies the microcausality condition.
As analyzed in Sec. 2, this is also the demand of physical
measurements. Therefore we need to calculate the function
$$
  A(x,y)=\langle\Psi\vert[:\varphi(x)\star\varphi(x):,
         :\varphi(y)\star\varphi(y):]_{\star}\vert\Psi\rangle ~,
  \eqno{(3.8)}  $$
where $\vert\Psi\rangle$ is a state vector of scalar field quantum
system. In Eq. (3.8), we have adopted the normal orderings for the
field operators $\varphi(x)\star\varphi(x)$ and
$\varphi(y)\star\varphi(y)$. This means that an infinite
vacuum energy has been eliminated in the corresponding commutative
spacetime field theory. To be the limit of physical measurements, we
take the state vector $\vert\Psi\rangle$ in Eq. (3.8) to be the vacuum
state $\vert0\rangle$. Therefore in this section, we will first
calculate the function
$$
  A_{0}(x,y)=\langle0\vert[:\varphi(x)\star\varphi(x):,
         :\varphi(y)\star\varphi(y):]_{\star}\vert0\rangle ~.
  \eqno{(3.9)}  $$
For the non-vacuum state expectation value of Eq. (3.8), we will
analyze it in Sec. 4.

  We decompose $\varphi(x)$ into the creation (negative frequency)
and annihilation (positive frequency) part
$$
  \varphi(x)=\varphi^{(-)}(x)+\varphi^{(+)}(x) ~,
  \eqno{(3.10)}  $$
where
$$
  \varphi^{-}(x)=\int\frac{d^{3}k}{\sqrt{(2\pi)^{3}2\omega_{k}}}
             a^{\dagger}(k)e^{-i{\bf k}\cdot{\bf x}+i\omega t}
            =\int\frac{d^{3}k}{\sqrt{(2\pi)^{3}2\omega_{k}}}
             a^{\dagger}(k)e^{ikx}
  \eqno{(3.11)}  $$
and
$$
  \varphi^{+}(x)=\int\frac{d^{3}k}{\sqrt{(2\pi)^{3}2\omega_{k}}}
           a(k)e^{i{\bf k}\cdot{\bf x}-i\omega t}
         =\int\frac{d^{3}k}{\sqrt{(2\pi)^{3}2\omega_{k}}}
           a(k)e^{-ikx} ~.
  \eqno{(3.12)}  $$
Here we define $kx=k_{\mu}x^{\mu}$. From Eq. (3.10) we have
$$
  \varphi(x)\star\varphi(x)=\varphi^{(-)}(x)\star\varphi^{(-)}(x)+
       \varphi^{(+)}(x)\star\varphi^{(-)}(x)+
       \varphi^{(-)}(x)\star\varphi^{(+)}(x)+
       \varphi^{(+)}(x)\star\varphi^{(+)}(x) ~.
  \eqno{(3.13)}  $$
The normal ordering of the operator $\varphi(x)\star\varphi(x)$ is
given by
$$
  :\varphi(x)\star\varphi(x):=\varphi^{(-)}(x)\star
        \varphi^{(-)}(x)+2\varphi^{(-)}(x)\star\varphi^{(+)}(x)+
        \varphi^{(+)}(x)\star\varphi^{(+)}(x) ~.
  \eqno{(3.14)}  $$
Here we have made a simplified manipulation for the normal ordering
of the Moyal star-product operator
$\varphi^{(+)}(x)\star\varphi^{(-)}(x)$. This is because the result
of the Moyal star-product between two functions is related with the
order of the two functions. In the Fourier integral representation,
we can see that $\varphi^{(-)}(x)\star\varphi^{(+)}(x)$ will have an
additional phase factor $e^{ik\times k^{\prime}}$ relative to
$\varphi^{(+)}(x)\star\varphi^{(-)}(x)$. However in Eq. (3.14) we
have ignored such a difference. The reason is that the terms that
contain $\varphi^{(-)}(x)\star\varphi^{(+)}(x)$ in the expansion of
Eq. (3.9) will contribute zero when we evaluate the vacuum
expectation values, as can be seen in the following. Thus we can
ignore such a difference equivalently for convenience.

  To expand $:\varphi(x)\star\varphi(x):\star:\varphi(y)
\star\varphi(y):$, we obtain
\begin{eqnarray*}
  & ~ & :\varphi(x)\star\varphi(x):\star:\varphi(y)\star\varphi(y): \\
  & = & \varphi^{(-)}(x)\star\varphi^{(-)}(x)\star\varphi^{(-)}(y)
        \star\varphi^{(-)}(y)+\varphi^{(-)}(x)\star\varphi^{(-)}(x)
        \star2\varphi^{(-)}(y)\star\varphi^{(+)}(y)       \\
  & ~ & +\varphi^{(-)}(x)\star\varphi^{(-)}(x)\star\varphi^{(+)}(y)
         \star\varphi^{(+)}(y)+2\varphi^{(-)}(x)\star\varphi^{(+)}(x)
         \star2\varphi^{(-)}(y)\star\varphi^{(-)}(y)      \\
  & ~ & +2\varphi^{(-)}(x)\star\varphi^{(+)}(x)\star2\varphi^{(-)}(y)
         \star\varphi^{(+)}(y)+2\varphi^{(-)}(x)\star\varphi^{(+)}(x)
         \star\varphi^{(+)}(y)\star\varphi^{(+)}(y)       \\
  & ~ & +\varphi^{(+)}(x)\star\varphi^{(+)}(x)\star\varphi^{(-)}(y)
         \star\varphi^{(-)}(y)+\varphi^{(+)}(x)\star\varphi^{(+)}(x)
         \star2\varphi^{(-)}(y)\star\varphi^{(+)}(y)      \\
  & ~ & +\varphi^{(+)}(x)\star\varphi^{(+)}(x)\star\varphi^{(+)}(y)
         \star\varphi^{(+)}(y) ~.
~~~~~~~~~~~~~~~~~~~~~~~~~~~~~~~~~~~~~~~~~~~~~~~~~(3.15)
\end{eqnarray*}
From Eq. (3.15), we can see clearly that the non-zero contributions
to the vacuum expectation value of
$:\varphi(x)\star\varphi(x):\star:\varphi(y)\star\varphi(y):$ come
from the terms which the most right hand side components of the
product operators are the negative frequency operators, and at the
same time for these terms the number of the positive frequency
component operators are equal to the number of the negative
frequency component operators in the total product operators.
Therefore we can see that there is only one such term
$\varphi^{(+)}(x)\star\varphi^{(+)}(x)\star\varphi^{(-)}(y)\star
 \varphi^{(-)}(y)$ which will contribute to non-zero vacuum
expectation value. Thus we have
$$
  \langle0\vert:\varphi(x)\star\varphi(x):\star:\varphi(y)
    \star\varphi(y):\vert0\rangle = \langle0\vert\varphi^{(+)}(x)
    \star\varphi^{(+)}(x)\star\varphi^{(-)}(y)\star\varphi^{(-)}(y)
    \vert0\rangle ~.
  \eqno{(3.16)}  $$
Similarly, the non-zero contribution to the vacuum expectation
value of
$:\varphi(y)\star\varphi(y):\star:\varphi(x)\star\varphi(x):$
only comes from the part $\varphi^{(+)}(y)\star\varphi^{(+)}(y)
\star\varphi^{(-)}(x)\star\varphi^{(-)}(x)$. We have
$$
  \langle0\vert:\varphi(y)\star\varphi(y):\star:\varphi(x)
    \star\varphi(x):\vert0\rangle = \langle0\vert\varphi^{(+)}(y)
    \star\varphi^{(+)}(y)\star\varphi^{(-)}(x)\star\varphi^{(-)}(x)
    \vert0\rangle ~.
  \eqno{(3.17)}  $$
If we do not use the normal orderings for the operators
$\varphi(x)\star\varphi(x)$ and $\varphi(y)\star\varphi(y)$ as that
of Eq. (3.7), then in the calculation of the vacuum expectation value
for Eq. (3.7), we need to consider the additional four terms which will
contribute non-zero results
$$
  \varphi^{(+)}(x)\star\varphi^{(-)}(x)\star\varphi^{(+)}(y)
    \star\varphi^{(-)}(y)-\varphi^{(+)}(y)\star\varphi^{(-)}(y)\star
    \varphi^{(+)}(x)\star\varphi^{(-)}(x) ~,         $$
$$
  \varphi^{(-)}(x)\star\varphi^{(+)}(x)\star\varphi^{(+)}(y)
    \star\varphi^{(-)}(y)-\varphi^{(-)}(y)\star\varphi^{(+)}(y)\star
    \varphi^{(+)}(x)\star\varphi^{(-)}(x) ~.         $$
However we can obtain that the total vacuum expectation value of
such four terms cancel at last. Thus to take the normal orderings
for the operators $\varphi(x)\star\varphi(x)$ and
$\varphi(y)\star\varphi(y)$ has simplified the calculation.

  Through calculation we obtain
\begin{eqnarray*}
  & ~ & \langle0\vert[:\varphi(x)\star\varphi(x):,
         :\varphi(y)\star\varphi(y):]_{\star}\vert0\rangle       \\
  & = &  \langle0\vert\varphi^{(+)}(x)\star\varphi^{(+)}(x)\star
         \varphi^{(-)}(y)\star\varphi^{(-)}(y)\vert0\rangle
        -\langle0\vert\varphi^{(+)}(y)\star\varphi^{(+)}(y)\star
         \varphi^{(-)}(x)\star\varphi^{(-)}(x)\vert0\rangle      \\
  & = & \int\frac{d^{3}k_{1}}{(2\pi)^{3}2\omega_{1}}
        \int\frac{d^{3}k_{2}}{(2\pi)^{3}2\omega_{2}}
        \Big[e^{-ik_{2}x}\star e^{-ik_{1}x}\star e^{ik_{2}y}
        \star e^{ik_{1}y}+e^{-ik_{1}x}
        \star e^{-ik_{2}x}\star e^{ik_{2}y}\star e^{ik_{1}y}     \\
  & ~ & ~~~~~~~~~~~~~~~~~~~~~~~~~
       -e^{-ik_{2}y}\star e^{-ik_{1}y}\star e^{ik_{2}x}\star
        e^{ik_{1}x}-e^{-ik_{1}y}\star e^{-ik_{2}y}\star e^{ik_{2}x}
          \star e^{ik_{1}x}\Big]                 \\
  & = & \int\frac{d^{3}k_{1}}{(2\pi)^{3}2\omega_{1}}
        \int\frac{d^{3}k_{2}}{(2\pi)^{3}2\omega_{2}}
        \Big[e^{-ik_{2}\times k_{1}}+1\Big]
        \Big[e^{-i(k_{2}+k_{1})x+i(k_{2}+k_{1})y}-
             e^{-i(k_{2}+k_{1})y+i(k_{2}+k_{1})x}\Big]    \\
  & = & \int\frac{d^{3}k_{1}}{(2\pi)^{3}2\omega_{1}}
        \int\frac{d^{3}k_{2}}{(2\pi)^{3}2\omega_{2}}(-2i)
        \Big[1+e^{ik_{1}\times k_{2}}\Big]\sin(k_{1}+k_{2})(x-y) ~.
~~~~~~~~~~~~~~~~~~~~~~~~(3.18)
\end{eqnarray*}
In Eq. (3.18), the first term of the third line means that two
scalar field quanta $\vert k_{1}\rangle$ and $\vert k_{2}\rangle$
are generated at spacetime point $y$ and annihilated at spacetime
point $x$. Because Moyal star-products depend on the orders of the
functions, there is the second term of the third line that is
responsible to the first term of the second line. Similarly, the two
terms of the fourth line mean that two scalar field quanta $\vert
k_{1}\rangle$ and $\vert k_{2}\rangle$ are generated at spacetime
point $x$ and annihilated at spacetime point $y$. In Eq. (3.18),
$\int\frac{d^{3}k_{1}}{(2\pi)^{3}2\omega_{1}}
\int\frac{d^{3}k_{2}}{(2\pi)^{3}2\omega_{2}}$ is the Lorentz
invariant volume element, $k_{1}\times
k_{2}=k_{1\mu}\theta^{\mu\nu}k_{2\nu}$, and
$(k_{1}+k_{2})(x-y)=(k_{1}+k_{2})_{\mu}(x-y)^{\mu}$. The total
expression is Lorentz invariant if we suppose that $\theta^{\mu\nu}$
is a second-order antisymmetric tensor. In the above calculation, we
have used Eq. (1.5) for the Moyal star-product of two functions
defined at two different spacetime points.

  We need to analyze whether the expression of Eq. (3.18)
disappears for a spacelike interval. This can be seen through the
vacuum expectation value of the equal-time commutator. Thus to take
$x_{0}=y_{0}$ in Eq. (3.18), we have
\begin{eqnarray*}
~~~~~~~~~~~~
  & ~ & \langle0\vert[:\varphi({\bf x},t)\star\varphi({\bf x},t):,
         :\varphi({\bf y},t)\star\varphi({\bf y},t):]_{\star}
          \vert0\rangle     \\
  & = & \int\frac{d^{3}k_{1}}{(2\pi)^{3}2\omega_{1}}
        \int\frac{d^{3}k_{2}}{(2\pi)^{3}2\omega_{2}}(2i)
        \Big[1+e^{ik_{1}\times k_{2}}\Big]\sin({\bf k}_{1}+
        {\bf k}_{2})\cdot({\bf x}-{\bf y}) ~.
~~~~~~~~~~~(3.19)
\end{eqnarray*}
We expand $k_{1}\times k_{2}$ as
$$
  k_{1}\times k_{2}=(k_{10},-{\bf k}_{1})\times(k_{20},-{\bf k}_{2})
        =k_{1i}\theta^{ij}k_{2j}-k_{1i}
         \theta^{i0}k_{20}-k_{10}\theta^{0i}k_{2i} ~.
  \eqno{(3.20)}  $$
For the case $\theta^{0i}=0$ of the spacetime noncommutativity, we
have
\begin{eqnarray*}
~~~~~~~~~~~~
  & ~ & \langle0\vert[:\varphi({\bf x},t)\star\varphi({\bf x},t):,
         :\varphi({\bf y},t)\star\varphi({\bf y},t):]_{\star}
          \vert0\rangle             \\
  & = & \int\frac{d^{3}k_{1}}{(2\pi)^{3}2\omega_{1}}
        \int\frac{d^{3}k_{2}}{(2\pi)^{3}2\omega_{2}}(2i)
        \Big[1+e^{ik_{1i}\theta^{ij}k_{2j}}\Big]\sin({\bf k}_{1}+
         {\bf k}_{2})\cdot({\bf x}-{\bf y}) ~.
~~~~~~~~(3.21)
\end{eqnarray*}
We can see that in Eq. (3.21), the integrand is a odd function to
the arguments (${\bf k}_{1}$,${\bf k}_{2}$). The integrand changes
its sign when the arguments (${\bf k}_{1}$,${\bf k}_{2}$) change to
($-{\bf k}_{1}$,$-{\bf k}_{2}$), while the integral measure does not
change. The integral space is symmetrical to the integral arguments
(${\bf k}_{1}$,${\bf k}_{2}$) and ($-{\bf k}_{1}$,$-{\bf k}_{2}$).
This makes the total integral of Eq. (3.21) to be zero. We have seen
that the total expression of Eq. (3.18) is Lorentz invariant in the
sense $\theta^{\mu\nu}$ being a second-order antisymmetric tensor.
This means that for an arbitrary spacelike interval of $x$ and $y$,
the integral of Eq. (3.18) vanishes. Thus we have
$$
  A_{0}(x,y)=\langle0\vert[:\varphi(x)\star\varphi(x):,
         :\varphi(y)\star\varphi(y):]_{\star}\vert0\rangle=0
  ~~~~~~ \mbox{for} ~~~~ (x-y)^{2}<0
  ~~~~ \mbox{when} ~~~~ \theta^{0i}=0 ~.
  \eqno{(3.22)}  $$
Therefore microcausality for the quadratic operator
$:\varphi(x)\star\varphi(x):$ of free scalar field is maintained for
the case $\theta^{0i}=0$ of spacetime noncommutativity.

  For the case $\theta^{0i}\neq0$ of spacetime noncommutativity,
we write $e^{ik_{1}\times k_{2}}$ as
\begin{eqnarray*}
~~~~~~~~~~~~
  & ~ & e^{ik_{1}\times k_{2}} = e^{i(k_{1i}\theta^{ij}k_{2j}-
           k_{1i}\theta^{i0}k_{20}-k_{10}\theta^{0i}k_{2i})}     \\
  & = & e^{ik_{1i}\theta^{ij}k_{2j}}\left[\cos(k_{1i}
         \theta^{i0}k_{20}+k_{10}\theta^{0i}k_{2i})-i\sin(k_{1i}
         \theta^{i0}k_{20}+k_{10}\theta^{0i}k_{2i})\right] ~.
~~~~~~~(3.23)
\end{eqnarray*}
From Eq. (3.19), we have
\begin{eqnarray*}
~
  & ~ & \langle0\vert[:\varphi({\bf x},t)\star\varphi({\bf x},t):,
         :\varphi({\bf y},t)\star\varphi({\bf y},t):]_{\star}
          \vert0\rangle     \\
  & = & \int\frac{d^{3}k_{1}}{(2\pi)^{3}2\omega_{1}}
        \int\frac{d^{3}k_{2}}{(2\pi)^{3}2\omega_{2}}(2i)
        \Big[1+e^{ik_{1i}\theta^{ij}k_{2j}}\cos(k_{1i}
         \theta^{i0}k_{20}+k_{10}\theta^{0i}k_{2i})         \\
  & ~ & ~~~~~~~~~~~~~~~~~~~ -ie^{ik_{1i}\theta^{ij}k_{2j}}\sin(k_{1i}
         \theta^{i0}k_{20}+k_{10}\theta^{0i}k_{2i})\Big]
         \sin({\bf k}_{1}+{\bf k}_{2})\cdot({\bf x}-{\bf y}) ~.
~~~~~~~(3.24)
\end{eqnarray*}
In the above integral, the integrand has three parts. The first two
parts are odd functions to the arguments (${\bf k}_{1}$,${\bf
k}_{2}$). While the integral measure does not change when the
arguments (${\bf k}_{1}$,${\bf k}_{2}$) change to ($-{\bf
k}_{1}$,$-{\bf k}_{2}$). The integral space is symmetrical to
integral arguments (${\bf k}_{1}$,${\bf k}_{2}$) and ($-{\bf
k}_{1}$,$-{\bf k}_{2}$). Hence the contribution of the first two
parts to the integral vanishes. The third part of the integrand is
an even function, its contribution to the integral does not vanish.
Thus we have
\begin{eqnarray*}
  & ~ & \langle0\vert[:\varphi({\bf x},t)\star\varphi({\bf x},t):,
         :\varphi({\bf y},t)\star\varphi({\bf y},t):]_{\star}
          \vert0\rangle             \\
  & = & \int\frac{d^{3}k_{1}}{(2\pi)^{3}2\omega_{1}}
        \int\frac{d^{3}k_{2}}{(2\pi)^{3}2\omega_{2}}
         2e^{ik_{1i}\theta^{ij}k_{2j}}
        \sin(k_{1i}\theta^{i0}k_{20}+k_{10}\theta^{0i}k_{2i})
        \sin({\bf k}_{1}+{\bf k}_{2})\cdot({\bf x}-{\bf y}) ~.
(3.25)
\end{eqnarray*}
Because the total expression of Eq. (3.18) is Lorentz invariant, we
have
$$
  A_{0}(x,y)=\langle0\vert[:\varphi(x)\star\varphi(x):,
         :\varphi(y)\star\varphi(y):]_{\star}\vert0\rangle\neq0
  ~~~~~~ \mbox{for} ~~~~ (x-y)^{2}<0
  ~~~~ \mbox{when} ~~~~ \theta^{0i}\neq0 ~.
  \eqno{(3.26)}  $$
This means that microcausality is violated for the quadratic
operator $:\varphi(x)\star\varphi(x):$ of the free scalar field for
the case $\theta^{0i}\neq0$ of spacetime noncommutativity. In the
limit $\theta^{0i}=0$, the integral of Eq. (3.25) vanishes and hence
$A_{0}(x,y)$ vanishes for $(x-y)^{2}<0$, which is in accordance with
Eq. (3.22).

\section{Non-vacuum state expectation values}

\indent

  In Sec. 3, we have calculated the expectation value $A_{0}(x,y)$
for the Moyal commutator of the quadratic operator
$:\varphi(x)\star\varphi(x):$. In this section we will analyze the
non-vacuum state expectation value $A(x,y)$ for the Moyal commutator
of the quadratic operator $:\varphi(x)\star\varphi(x):$. As defined
in Eq. (3.8), we write
$$
  A(x,y)=\langle\Psi\vert[:\varphi(x)\star\varphi(x):,
         :\varphi(y)\star\varphi(y):]_{\star}\vert\Psi\rangle ~.
  \eqno{(4.1)}  $$
In order to evaluate Eq. (4.1), we first need to define the state
vector $\vert\Psi\rangle$ for a scalar field quantum system.

  Supposing that the state vector $\vert\Psi\rangle$ is in the
occupation eigenstate, we can write
$$
  \vert\Psi\rangle=\vert N_{k_{1}}N_{k_{2}}\cdots N_{k_{i}}\cdots,0
              \rangle ~,
  \eqno{(4.2)}  $$
where $N_{k_{i}}$ represents the occupation number of the momentum
$k_{i}$. For an arbitrary actual field quantum system, the total
energy is always finite. Because the occupation numbers $N_{k_{i}}$
take values of finitely integral numbers, the occupation numbers
$N_{k_{i}}$ should only be nonzero on finite number separate
momentums $k_{i}$. Otherwise, if $N_{k_{i}}$ take nonzero values on
infinite number separate momentums $k_{i}$, or on a continuous
interval of the momentum, the total energy of the scalar field
quantum system will be infinite. While for a actual field quantum
system, the total energy is always finite. These properties for the
state vector $\vert\Psi\rangle$ are very useful for the following
calculations. In Eq. (4.2), we use $0$ to represent that the
occupation numbers are all zero on the other momentums except for
$k_{i}$. The state vector $\vert\Psi\rangle$ has the following
properties [27]:
$$
  \langle N_{k_{1}}N_{k_{2}}\cdots N_{k_{i}}\cdots\vert
          N_{k_{1}}N_{k_{2}}\cdots N_{k_{i}}\cdots\rangle =1 ~,
  \eqno{(4.3)}  $$
$$
  \sum\limits_{N_{k_{1}}N_{k_{2}}\cdots}
    \vert N_{k_{1}}N_{k_{2}}\cdots N_{k_{i}}\cdots\rangle
    \langle N_{k_{1}}N_{k_{2}}\cdots N_{k_{i}}\cdots\vert=1 ~,
  \eqno{(4.4)}  $$
$$
  a(k_{i})\vert N_{k_{1}}N_{k_{2}}\cdots N_{k_{i}}\cdots\rangle=
          \sqrt{N_{k_{i}}}\vert N_{k_{1}}N_{k_{2}}\cdots
                (N_{k_{i}}-1)\cdots\rangle ~,
  \eqno{(4.5)}  $$
$$
  a^{\dagger}(k_{i})\vert N_{k_{1}}N_{k_{2}}\cdots N_{k_{i}}
          \cdots\rangle=\sqrt{N_{k_{i}}+1}\vert N_{k_{1}}N_{k_{2}}
          \cdots(N_{k_{i}}+1)\cdots\rangle ~.
  \eqno{(4.6)}  $$
In Eqs. (4.3)-(4.6), we have omitted the notation $0$ of Eq. (4.2)
in the state vectors for convenience. Equation (4.4) is the
completeness expression. Thus Eq. (4.2) can represent an arbitrary
scalar field quantum system.

  In Ref. [22], we have obtained that the non-vacuum state expectation
values for the Moyal commutator $[\varphi(x),\varphi(y)]_{\star}$ of
the scalar field and Moyal anticommutator
$\{\psi_{\alpha}(x),\overline{\psi}_{\beta}(x^{\prime})\}_{\star}$
of the Dirac field are just equal to their vacuum state expectation
values. If such a property is still held for the quadratic operator
$:\varphi(x)\star\varphi(x):$ of scalar field studied in this paper,
we can obtain a direct answer for the function $A(x,y)$ of Eq.
(4.1). As can be seen in the following, such a property is really
held for the quadratic operator $:\varphi(x)\star\varphi(x):$.
However we need to verify this point.

  For the convenience of the analyzing, we decompose the state vector
of Eq. (4.2) into two parts
$$
  \vert\Psi\rangle=\vert(k_{1})(k_{2})\cdots(k_{i})\cdots0\rangle+
    \vert N_{k_{1}}N_{k_{2}}\cdots N_{k_{i}}\cdots\rangle ~.
  \eqno{(4.7)}  $$
In Eq. (4.7), we use
$\vert(k_{1})(k_{2})\cdots(k_{i})\cdots0\rangle$ to represent that
the state is on the vacuum, while the finite number separate
momentums $k_{i}$ are eliminated from the arguments of $k$ for such
a vacuum state. And we use $\vert N_{k_{1}}N_{k_{2}}\cdots
N_{k_{i}}\cdots\rangle$ to represent a non-vacuum state which the
arguments only take the finite number separate values $k_{i}$. On
these separate $k_{i}$, the occupation numbers are $N_{k_{i}}$,
which take values of finite integrals. Then for Eq. (4.1), we can
write
\begin{eqnarray*}
  & ~ & \langle\Psi\vert[:\varphi(x)\star\varphi(x):,
         :\varphi(y)\star\varphi(y):]_{\star}\vert\Psi\rangle     \\
  & = & ~~~ \langle(k_{1})(k_{2})\cdots(k_{i})\cdots0\vert
         [:\varphi(x)\star\varphi(x):,
          :\varphi(y)\star\varphi(y):]_{\star}
        \vert(k_{1})(k_{2})\cdots(k_{i})\cdots0\rangle            \\
  & ~ & + \langle N_{k_{1}}N_{k_{2}}\cdots N_{k_{i}}\cdots\vert
         [:\varphi(x)\star\varphi(x):,
          :\varphi(y)\star\varphi(y):]_{\star}
          \vert N_{k_{1}}N_{k_{2}}\cdots N_{k_{i}}\cdots\rangle   \\
  & ~ & + \langle(k_{1})(k_{2})\cdots(k_{i})\cdots0\vert
         [:\varphi(x)\star\varphi(x):,
          :\varphi(y)\star\varphi(y):]_{\star}
          \vert N_{k_{1}}N_{k_{2}}\cdots N_{k_{i}}\cdots\rangle   \\
  & ~ & + \langle N_{k_{1}}N_{k_{2}}\cdots N_{k_{i}}\cdots\vert
         [:\varphi(x)\star\varphi(x):,
          :\varphi(y)\star\varphi(y):]_{\star}
        \vert(k_{1})(k_{2})\cdots(k_{i})\cdots0\rangle ~.
~~~~~(4.8)
\end{eqnarray*}
We can see that the last two terms of Eq. (4.8) are all zero,
because their arguments of the momentum $k$ will not match with each
other for the bras and kets. Hence we have
\begin{eqnarray*}
~~~
  & ~ & \langle\Psi\vert[:\varphi(x)\star\varphi(x):,
         :\varphi(y)\star\varphi(y):]_{\star}\vert\Psi\rangle   \\
  & = & ~~~ \langle(k_{1})(k_{2})\cdots(k_{i})\cdots0\vert
         [:\varphi(x)\star\varphi(x):,
          :\varphi(y)\star\varphi(y):]_{\star}
        \vert(k_{1})(k_{2})\cdots(k_{i})\cdots0\rangle          \\
  & ~ & + \langle N_{k_{1}}N_{k_{2}}\cdots N_{k_{i}}\cdots\vert
         [:\varphi(x)\star\varphi(x):,
          :\varphi(y)\star\varphi(y):]_{\star}
          \vert N_{k_{1}}N_{k_{2}}\cdots N_{k_{i}}\cdots\rangle ~.
~~~~~~~~(4.9)
\end{eqnarray*}

  For the first term of Eq. (4.9), its calculation is just like that
of Eq. (3.18), except that the separate momentums $k_{i}$ should be
eliminated from the final integral measure. Thus according to the
result of Eq. (3.18), we obtain
\begin{eqnarray*}
  & ~ & \langle(k_{1})(k_{2})\cdots(k_{i})\cdots0\vert
         [:\varphi(x)\star\varphi(x):,
          :\varphi(y)\star\varphi(y):]_{\star}
        \vert(k_{1})(k_{2})\cdots(k_{i})\cdots0\rangle          \\
  & = & \int\limits_{(\mbox{$k_{i}$ eliminated})}
        \frac{d^{3}k_{a}}{(2\pi)^{3}2\omega_{a}}
        \int\limits_{(\mbox{$k_{i}$ eliminated})}
        \frac{d^{3}k_{b}}{(2\pi)^{3}2\omega_{b}}(-2i)
        \Big[1+e^{ik_{a}\times k_{b}}\Big]\sin(k_{a}+k_{b})(x-y) ~.
\end{eqnarray*}
$$  \eqno{(4.10)}  $$
In Eq. (4.10), we use ($k_{i}$ eliminated) to represent that
in the integral for $k_{a}$ and $k_{b}$, a set of finite number
separate points $k_{i}$ is eliminated from the total integral measure
of $k_{a}$ and $k_{b}$ respectively. We can write Eq. (4.10)
equivalently in the form
\begin{eqnarray*}
  & ~ & \langle(k_{1})(k_{2})\cdots(k_{i})\cdots0\vert
         [:\varphi(x)\star\varphi(x):,
          :\varphi(y)\star\varphi(y):]_{\star}
        \vert(k_{1})(k_{2})\cdots(k_{i})\cdots0\rangle          \\
  & = & \int\frac{d^{3}k_{a}}{(2\pi)^{3}2\omega_{a}}
        \int\frac{d^{3}k_{b}}{(2\pi)^{3}2\omega_{b}}(-2i)
        \Big[1+e^{ik_{a}\times k_{b}}\Big]\sin(k_{a}+k_{b})(x-y)   \\
  & ~ & -\int\limits_{(\mbox{only on $k_{i}$})}
        \frac{d^{3}k_{a}}{(2\pi)^{3}2\omega_{a}}
        \int\limits_{(\mbox{only on $k_{i}$})}
        \frac{d^{3}k_{b}}{(2\pi)^{3}2\omega_{b}}(-2i)
        \Big[1+e^{ik_{a}\times k_{b}}\Big]\sin(k_{a}+k_{b})(x-y) ~.
\end{eqnarray*}
$$  \eqno{(4.11)}  $$
In Eq. (4.11), we use (only on $k_{i}$) to represent that in the
second integral, the integral is only taken on a set of finite
number separate points $k_{i}$ for $k_{a}$ and $k_{b}$ respectively.
Because the integrand is a bounded function, while the integral
measure is zero for the second integral, according to the theory of
integration (for example see Ref. [28]), we obtain that the second
part of Eq. (4.11) is zero. Therefore we obtain
\begin{eqnarray*}
~~~~
  & ~ & \langle(k_{1})(k_{2})\cdots(k_{i})\cdots0\vert
         [:\varphi(x)\star\varphi(x):,
          :\varphi(y)\star\varphi(y):]_{\star}
        \vert(k_{1})(k_{2})\cdots(k_{i})\cdots0\rangle          \\
  & = & \int\frac{d^{3}k_{a}}{(2\pi)^{3}2\omega_{a}}
        \int\frac{d^{3}k_{b}}{(2\pi)^{3}2\omega_{b}}(-2i)
        \Big[1+e^{ik_{a}\times k_{b}}\Big]\sin(k_{a}+k_{b})(x-y) ~.
~~~~~~~~~~~~~~~~~(4.12)
\end{eqnarray*}
Or we can write
\begin{eqnarray*}
~~
  & ~ & \langle(k_{1})(k_{2})\cdots(k_{i})\cdots0\vert
         [:\varphi(x)\star\varphi(x):,
          :\varphi(y)\star\varphi(y):]_{\star}
        \vert(k_{1})(k_{2})\cdots(k_{i})\cdots0\rangle          \\
  & = & \langle0\vert[:\varphi(x)\star\varphi(x):,
         :\varphi(y)\star\varphi(y):]_{\star}\vert0\rangle
          =A_{0}(x,y) ~.
~~~~~~~~~~~~~~~~~~~~~~~~~~~~~~~~~~(4.13)
\end{eqnarray*}

  For the second term of (4.9), we have
\begin{eqnarray*}
~~~
  & ~ &  \langle N_{k_{1}}N_{k_{2}}\cdots N_{k_{i}}\cdots\vert
         [:\varphi(x)\star\varphi(x):,
          :\varphi(y)\star\varphi(y):]_{\star}
          \vert N_{k_{1}}N_{k_{2}}\cdots N_{k_{i}}\cdots\rangle    \\
  & = & ~~ \langle N_{k_{1}}N_{k_{2}}\cdots N_{k_{i}}\cdots\vert
         :\varphi(x)\star\varphi(x):\star :\varphi(y)\star\varphi(y):
          \vert N_{k_{1}}N_{k_{2}}\cdots N_{k_{i}}\cdots\rangle    \\
  & ~ &  -\langle N_{k_{1}}N_{k_{2}}\cdots N_{k_{i}}\cdots\vert
         :\varphi(y)\star\varphi(y):\star :\varphi(x)\star\varphi(x):
          \vert N_{k_{1}}N_{k_{2}}\cdots N_{k_{i}}\cdots\rangle ~.
~~~~~~(4.14)
\end{eqnarray*}
We first analyze the first term of Eq. (4.14). We can find from Eq.
(3.15) that the nonzero contributions not only come from the
operator $\varphi^{(+)}(x)\star\varphi^{(+)}(x)\star\varphi^{(-)}(y)
\star\varphi^{(-)}(y)$ like that of Eq. (3.16), but also come from
the operators
$\varphi^{(-)}(x)\star\varphi^{(-)}(x)\star\varphi^{(+)}(y)
\star\varphi^{(+)}(y)$ and
$2\varphi^{(-)}(x)\star\varphi^{(+)}(x)\star2\varphi^{(-)}(y)
\star\varphi^{(+)}(y)$ of Eq. (3.15) which have the equal numbers of
negative frequency and positive frequency components. The situation
for the second term of Eq. (4.14) is similar. For the total
integrand generated from these operators, we can note it as
$G(k_{a},k_{b},x,y)$. We can write Eq. (4.14) as
\begin{eqnarray*}
~~~~~
  & ~ & \langle N_{k_{1}}N_{k_{2}}\cdots N_{k_{i}}\cdots\vert
        [:\varphi(x)\star\varphi(x):,
         :\varphi(y)\star\varphi(y):]_{\star}
         \vert N_{k_{1}}N_{k_{2}}\cdots N_{k_{i}}\cdots\rangle     \\
  & = & \int\limits_{(\mbox{only on $k_{i}$})}
        \frac{d^{3}k_{a}}{(2\pi)^{3}2\omega_{a}}
        \int\limits_{(\mbox{only on $k_{i}$})}
        \frac{d^{3}k_{b}}{(2\pi)^{3}2\omega_{b}} ~
         G(k_{a},k_{b},x,y) ~.
~~~~~~~~~~~~~~~~~~~~~~~~(4.15)
\end{eqnarray*}
The integrand $G(k_{a},k_{b},x,y)$ is not equal to the integrand of
Eq. (4.12) generally. However we need not to obtain its explicit
form in fact. This is because its contribution to the integral of
Eq. (4.15) is zero. The reason is that the integral of
$G(k_{a},k_{b},x,y)$ is only taken on finite number separate points
of $k_{a}$ and $k_{b}$, their total integral measure is zero. While
the integrand $G(k_{a},k_{b},x,y)$ should be a bounded function.
This makes the total integral of Eq. (4.15) be zero according to the
theory of integration (for example see Ref. [28]). Therefore we
obtain
$$
  \langle N_{k_{1}}N_{k_{2}}\cdots N_{k_{i}}\cdots\vert
         [:\varphi(x)\star\varphi(x):,
          :\varphi(y)\star\varphi(y):]_{\star}
          \vert N_{k_{1}}N_{k_{2}}\cdots N_{k_{i}}\cdots\rangle=0 ~.
  \eqno{(4.16)}  $$

  To combine the results of Eqs. (4.12) and (4.16) together, we obtain
\begin{eqnarray*}
~~~~~~~~
  & ~ & \langle\Psi\vert[:\varphi(x)\star\varphi(x):,
         :\varphi(y)\star\varphi(y):]_{\star}\vert\Psi\rangle   \\
  & = & \int\frac{d^{3}k_{1}}{(2\pi)^{3}2\omega_{1}}
        \int\frac{d^{3}k_{2}}{(2\pi)^{3}2\omega_{2}}(-2i)
        \Big[1+e^{ik_{1}\times k_{2}}\Big]\sin(k_{1}+k_{2})(x-y) ~.
~~~~~~~~~~~~~~~(4.17)
\end{eqnarray*}
Or we can write
$$
  \langle\Psi\vert[:\varphi(x)\star\varphi(x):,
         :\varphi(y)\star\varphi(y):]_{\star}\vert\Psi\rangle =
  \langle0\vert[:\varphi(x)\star\varphi(x):,
         :\varphi(y)\star\varphi(y):]_{\star}\vert0\rangle ~.
  \eqno{(4.18)}  $$
Therefore we obtain $A(x,y)=A_{0}(x,y)$, which is a universal
function for an arbitrary state vector of Eq. (4.2). Thus for the
quadratic operator $:\varphi(x)\star\varphi(x):$ of free scalar
field, its microcausality under the measurements of non-vacuum
states is equivalent to the measurement of the vacuum state. Thus
from the result of Sec. 3, we have
$$
  A(x,y)=\langle\Psi\vert[:\varphi(x)\star\varphi(x):,
         :\varphi(y)\star\varphi(y):]_{\star}\vert\Psi\rangle=0
  ~~~~~~ \mbox{for} ~~~~ (x-y)^{2}<0
  ~~~~ \mbox{when} ~~~~ \theta^{0i}=0 ~,
  \eqno{(4.19)}  $$
which means that microcausality is maintained for the quadratic
operator $:\varphi(x)\star\varphi(x):$ of free scalar field for the
case $\theta^{0i}=0$ of spacetime noncommutativity. For the case
$\theta^{0i}\neq0$, for the equal-time commutator, like that of Eq.
(3.25), we have
\begin{eqnarray*}
  & ~ & \langle\Psi\vert[:\varphi({\bf x},t)\star\varphi({\bf x},t):,
         :\varphi({\bf y},t)\star\varphi({\bf y},t):]_{\star}
          \vert\Psi\rangle             \\
  & = & \int\frac{d^{3}k_{1}}{(2\pi)^{3}2\omega_{1}}
        \int\frac{d^{3}k_{2}}{(2\pi)^{3}2\omega_{2}}
         2e^{ik_{1i}\theta^{ij}k_{2j}}
        \sin(k_{1i}\theta^{i0}k_{20}+k_{10}\theta^{0i}k_{2i})
        \sin({\bf k}_{1}+{\bf k}_{2})\cdot({\bf x}-{\bf y}) ~.     \\
\end{eqnarray*}
$$  \eqno{(4.20)}  $$
It is not zero because the integrand is an even function. If we
suppose that $\theta^{\mu\nu}$ is a second-order antisymmetric
tensor, the total expression of Eq. (4.17) is Lorentz invariant, we
have
$$
  A(x,y)=\langle\Psi\vert[:\varphi(x)\star\varphi(x):,
         :\varphi(y)\star\varphi(y):]_{\star}\vert\Psi\rangle\neq0
  ~~~~~~ \mbox{for} ~~~~ (x-y)^{2}<0
  ~~~~ \mbox{when} ~~~~ \theta^{0i}\neq0 ~,
  \eqno{(4.21)}  $$
which means that microcausality is violated for the quadratic
operator $:\varphi(x)\star\varphi(x):$ of free scalar field for the
case $\theta^{0i}\neq0$ of spacetime noncommutativity. In Eq.
(4.20), $\langle\Psi\vert[:\varphi({\bf x},t)\star\varphi({\bf
x},t):, :\varphi({\bf y},t)\star\varphi({\bf
y},t):]_{\star}\vert\Psi\rangle$ is also a universal function for an
arbitrary state vector of Eq. (4.2). In the limit $\theta^{0i}=0$,
the integral of Eq. (4.20) vanishes and hence $A(x,y)$ vanishes for
$(x-y)^{2}<0$, which is in accordance with Eq. (4.19).

  To summarize, we have obtained in Sec. 3 and this section that
microcausality for the quadratic operator $\varphi({\bf
x},t)\star\varphi({\bf x},t)$ of free scalar field on noncommutative
spacetime is satisfied for the case $\theta^{0i}=0$ of spacetime
noncommutativity, and is violated for the case $\theta^{0i}\neq0$ of
spacetime noncommutativity.

\section{Some other quadratic operators of
                    scalar field on noncommutative spacetime}

\indent

  In this section, we analyze the microcausality of some other
quadratic operators of free scalar field on noncommutative
spacetime. These quadratic operators include $\pi({\bf
x},t)\star\pi({\bf x},t)$, $\partial_{i}\varphi({\bf
x},t)\star\partial_{i}\varphi({\bf x},t)$, and
$\partial_{i}\varphi({\bf x},t)\star\pi({\bf x},t)$, which are
composition parts of the energy-momentum density of Eqs. (3.2) and
(3.3). We first calculate their vacuum expectation values. To be
brief, some processes are omitted. From the result of Eq. (3.18), we
can obtain
\begin{eqnarray*}
~~~~~~~
  & ~ & \langle0\vert[:\pi(x)\star\pi(x):,
         :\pi(y)\star\pi(y):]_{\star}\vert0\rangle     \\
  & = & \int\frac{d^{3}k_{1}}{(2\pi)^{3}2\omega_{1}}
        \int\frac{d^{3}k_{2}}{(2\pi)^{3}2\omega_{2}}
         (-2i\omega_{1}^{2}\omega_{2}^{2})
        \Big[1+e^{ik_{1}\times k_{2}}\Big]\sin(k_{1}+k_{2})(x-y) ~,
~~~~~~~~~~(5.1)
\end{eqnarray*}
\begin{eqnarray*}
~~~~~
  & ~ & \langle0\vert[:\partial_{i}\varphi(x)\star\partial_{i}
         \varphi(x):,:\partial_{i}\varphi(y)\star
         \partial_{i}\varphi(y):]_{\star}\vert0\rangle           \\
  & = & \int\frac{d^{3}k_{1}}{(2\pi)^{3}2\omega_{1}}
        \int\frac{d^{3}k_{2}}{(2\pi)^{3}2\omega_{2}}
         (-2ik_{1i}^{2}k_{2i}^{2})
        \Big[1+e^{ik_{1}\times k_{2}}\Big]\sin(k_{1}+k_{2})(x-y) ~,
~~~~~~~~~~~~(5.2)
\end{eqnarray*}
\begin{eqnarray*}
~~
  & ~ & \langle0\vert[:\partial_{i}\varphi(x)\star\pi(x):,
         :\partial_{i}\varphi(y)\star\pi(y):]_{\star}\vert0\rangle   \\
  & = & \int\frac{d^{3}k_{1}}{(2\pi)^{3}2\omega_{1}}
        \int\frac{d^{3}k_{2}}{(2\pi)^{3}2\omega_{2}}
         (-2ik_{2i}\omega_{1})\Big[k_{1i}\omega_{2}+k_{2i}\omega_{1}
         e^{ik_{1}\times k_{2}}\Big]\sin(k_{1}+k_{2})(x-y) ~.
~~~~~(5.3)
\end{eqnarray*}
We can see that the differences of these integrals to Eq. (3.18) lie
in the coefficient factors, such as $\omega_{1}^{2}\omega_{2}^{2}$,
$k_{1i}^{2}k_{2i}^{2}$, $k_{1i}k_{2i}\omega_{1}\omega_{2}$, and
$k_{2i}^{2}\omega_{1}^{2}$. These coefficient factors make these
integrals be not Lorentz invariant functions. The reason lies in the
fact that these operators are not Lorentz invariant themselves.

  We can also obtain the vacuum expectation values of the equal-time
Moyal commutators for these operators respectively. They are given by
\begin{eqnarray*}
~~~~
  & ~ & \langle0\vert[:\pi({\bf x},t)\star\pi({\bf x},t):,
         :\pi({\bf y},t))\star\pi({\bf y},t):]_{\star}\vert0\rangle  \\
  & = & \int\frac{d^{3}k_{1}}{(2\pi)^{3}2\omega_{1}}
        \int\frac{d^{3}k_{2}}{(2\pi)^{3}2\omega_{2}}
         (2i\omega_{1}^{2}\omega_{2}^{2})
        \Big[1+e^{ik_{1}\times k_{2}}\Big]\sin({\bf k}_{1}+
        {\bf k}_{2})\cdot({\bf x}-{\bf y}) ~,
~~~~~~~~~~~~(5.4)
\end{eqnarray*}
\begin{eqnarray*}
~~~~
  & ~ & \langle0\vert[:\partial_{i}\varphi({\bf x},t)\star\partial_{i}
         \varphi({\bf x},t):,:\partial_{i}\varphi({\bf y},t))\star
         \partial_{i}\varphi({\bf y},t)):]_{\star}\vert0\rangle      \\
  & = & \int\frac{d^{3}k_{1}}{(2\pi)^{3}2\omega_{1}}
        \int\frac{d^{3}k_{2}}{(2\pi)^{3}2\omega_{2}}
         (2ik_{1i}^{2}k_{2i}^{2})
        \Big[1+e^{ik_{1}\times k_{2}}\Big]\sin({\bf k}_{1}+
        {\bf k}_{2})\cdot({\bf x}-{\bf y}) ~,
~~~~~~~~~~~(5.5)
\end{eqnarray*}
\begin{eqnarray*}
~~
  & ~ & \langle0\vert[:\partial_{i}\varphi({\bf x},t)
        \star\pi({\bf x},t):,:\partial_{i}\varphi({\bf y},t))
        \star\pi({\bf y},t)):]_{\star}\vert0\rangle                 \\
  & = & \int\frac{d^{3}k_{1}}{(2\pi)^{3}2\omega_{1}}
        \int\frac{d^{3}k_{2}}{(2\pi)^{3}2\omega_{2}}
         (2ik_{2i}\omega_{1})\Big[k_{1i}\omega_{2}+k_{2i}\omega_{1}
         e^{ik_{1}\times k_{2}}\Big]\sin({\bf k}_{1}+
        {\bf k}_{2})\cdot({\bf x}-{\bf y}) ~.
~~~(5.6)
\end{eqnarray*}
As in Sec. 3, we write $k_{1}\times k_{2}$ as
$$
  k_{1}\times k_{2}=(k_{10},-{\bf k}_{1})\times(k_{20},-{\bf k}_{2})
        =k_{1i}\theta^{ij}k_{2j}-k_{1i}
         \theta^{i0}k_{20}-k_{10}\theta^{0i}k_{2i} ~.
  \eqno{(5.7)}  $$
For the case $\theta^{0i}=0$ of spacetime noncommutativity, we have
\begin{eqnarray*}
~~~~~
  & ~ & \langle0\vert[:\pi({\bf x},t)\star\pi({\bf x},t):,
         :\pi({\bf y},t))\star\pi({\bf y},t):]_{\star}\vert0\rangle  \\
  & = & \int\frac{d^{3}k_{1}}{(2\pi)^{3}2\omega_{1}}
        \int\frac{d^{3}k_{2}}{(2\pi)^{3}2\omega_{2}}
         (2i\omega_{1}^{2}\omega_{2}^{2})
        \Big[1+e^{ik_{1i}\theta^{ij}k_{2j}}\Big]\sin({\bf k}_{1}+
        {\bf k}_{2})\cdot({\bf x}-{\bf y}) ~,
~~~~~~~~~(5.8)
\end{eqnarray*}
\begin{eqnarray*}
~~~~~
  & ~ & \langle0\vert[:\partial_{i}\varphi({\bf x},t)\star\partial_{i}
         \varphi({\bf x},t):,:\partial_{i}\varphi({\bf y},t))\star
         \partial_{i}\varphi({\bf y},t)):]_{\star}\vert0\rangle      \\
  & = & \int\frac{d^{3}k_{1}}{(2\pi)^{3}2\omega_{1}}
        \int\frac{d^{3}k_{2}}{(2\pi)^{3}2\omega_{2}}
         (2ik_{1i}^{2}k_{2i}^{2})
        \Big[1+e^{ik_{1i}\theta^{ij}k_{2j}}\Big]\sin({\bf k}_{1}+
        {\bf k}_{2})\cdot({\bf x}-{\bf y}) ~,
~~~~~~~~~(5.9)
\end{eqnarray*}
\begin{eqnarray*}
  & ~ & \langle0\vert[:\partial_{i}\varphi({\bf x},t)
        \star\pi({\bf x},t):,:\partial_{i}\varphi({\bf y},t))
        \star\pi({\bf y},t)):]_{\star}\vert0\rangle                 \\
  & = & \int\frac{d^{3}k_{1}}{(2\pi)^{3}2\omega_{1}}
        \int\frac{d^{3}k_{2}}{(2\pi)^{3}2\omega_{2}}
         (2ik_{2i}\omega_{1})\Big[k_{1i}\omega_{2}+k_{2i}\omega_{1}
         e^{ik_{1i}\theta^{ij}k_{2j}}\Big]\sin({\bf k}_{1}+
        {\bf k}_{2})\cdot({\bf x}-{\bf y}) ~.
~~(5.10)
\end{eqnarray*}
We can see that in Eqs. (5.8)-(5.10), the integrands are all odd
functions to the arguments (${\bf k}_{1}$,${\bf k}_{2}$). The
integrands change their signs when the arguments (${\bf
k}_{1}$,${\bf k}_{2}$) change to ($-{\bf k}_{1}$,$-{\bf k}_{2}$),
while the integral measures do not change. The integral spaces are
symmetrical to the integral arguments (${\bf k}_{1}$,${\bf k}_{2}$)
and ($-{\bf k}_{1}$,$-{\bf k}_{2}$). Thus the total integrals of
Eqs. (5.8)-(5.10) are all zero. Although the integrals of Eqs.
(5.1)-(5.3) are not Lorentz invariant functions, this does not mean
that microcausality will violate for an arbitrary spacelike interval
of $x$ and $y$, because the Lorentz un-invariance of these integrals
comes from the fact that the considered operators are not Lorentz
invariant themselves. Thus from these results we can draw conclusion
that microcausality is maintained for these quadratic operators for
the case $\theta^{0i}=0$ of spacetime noncommutativity.

  For the case $\theta^{0i}\neq0$ of spacetime noncommutativity,
we write $e^{ik_{1}\times k_{2}}$ as
\begin{eqnarray*}
~~~~~~~
  & ~ & e^{ik_{1}\times k_{2}} = e^{i(k_{1i}\theta^{ij}k_{2j}-
           k_{1i}\theta^{i0}k_{20}-k_{10}\theta^{0i}k_{2i})}       \\
  & = & e^{ik_{1i}\theta^{ij}k_{2j}}\left[\cos(k_{1i}
         \theta^{i0}k_{20}+k_{10}\theta^{0i}k_{2i})-i\sin(k_{1i}
         \theta^{i0}k_{20}+k_{10}\theta^{0i}k_{2i})\right] ~.
~~~~~~~~~~~~~(5.11)
\end{eqnarray*}
To substitute Eq. (5.11) in Eqs. (5.4)-(5.6), and to remove away the
odd function parts in the integrands which will contribute zero to
the whole integrals, we obtain
\begin{eqnarray*}
  & ~ & \langle0\vert[:\pi({\bf x},t)\star\pi({\bf x},t):,
         :\pi({\bf y},t))\star\pi({\bf y},t):]_{\star}\vert0\rangle  \\
  & = & \int\frac{d^{3}k_{1}}{(2\pi)^{3}2\omega_{1}}
        \int\frac{d^{3}k_{2}}{(2\pi)^{3}2\omega_{2}}
         (2\omega_{1}^{2}\omega_{2}^{2})
         e^{ik_{1i}\theta^{ij}k_{2j}}
        \sin(k_{1i}\theta^{i0}k_{20}+k_{10}\theta^{0i}k_{2i})
        \sin({\bf k}_{1}+{\bf k}_{2})\cdot({\bf x}-{\bf y}) ~,
\end{eqnarray*}
$$  \eqno{(5.12)}  $$
\begin{eqnarray*}
  & ~ & \langle0\vert[:\partial_{i}\varphi({\bf x},t)\star\partial_{i}
         \varphi({\bf x},t):,:\partial_{i}\varphi({\bf y},t))\star
         \partial_{i}\varphi({\bf y},t)):]_{\star}\vert0\rangle     \\
  & = & \int\frac{d^{3}k_{1}}{(2\pi)^{3}2\omega_{1}}
        \int\frac{d^{3}k_{2}}{(2\pi)^{3}2\omega_{2}}
         (2k_{1i}^{2}k_{2i}^{2})
        e^{ik_{1i}\theta^{ij}k_{2j}}
        \sin(k_{1i}\theta^{i0}k_{20}+k_{10}
        \theta^{0i}k_{2i})\sin({\bf k}_{1}+
        {\bf k}_{2})\cdot({\bf x}-{\bf y}) ~,
\end{eqnarray*}
$$  \eqno{(5.13)}  $$
\begin{eqnarray*}
  & ~ & \langle0\vert[:\partial_{i}\varphi({\bf x},t)
        \star\pi({\bf x},t):,:\partial_{i}\varphi({\bf y},t))
        \star\pi({\bf y},t)):]_{\star}\vert0\rangle          \\
  & = & \int\frac{d^{3}k_{1}}{(2\pi)^{3}2\omega_{1}}
        \int\frac{d^{3}k_{2}}{(2\pi)^{3}2\omega_{2}}
         (2k_{2i}^{2}\omega_{1}^{2})e^{ik_{1i}
          \theta^{ij}k_{2j}}\sin(k_{1i}\theta^{i0}k_{20}+
           k_{10}\theta^{0i}k_{2i})\sin({\bf k}_{1}+
          {\bf k}_{2})\cdot({\bf x}-{\bf y}) ~.
\end{eqnarray*}
$$  \eqno{(5.14)}  $$
In Eqs. (5.12)-(5.14), the integrands are all even functions. This
makes the whole integrals not vanished. As pointed out above, the
integrals of Eqs. (5.1)-(5.3) are not Lorentz invariant functions,
while the Lorentz un-invariance of these integrals comes from the
fact that the considered operators are not Lorentz invariant
themselves. Thus for an arbitrary spacelike interval of $x$ and $y$,
equations (5.1)-(5.3) do not vanish either generally when
$\theta^{0i}\neq0$. This means that microcausality is violated for
these quadratic operators for the case $\theta^{0i}\neq0$ of
spacetime noncommutativity. In the limit $\theta^{0i}=0$, the
integrals of Eqs. (5.12)-(5.14) vanish, which is in accordance with
Eqs. (5.8)-(5.10). To be complete, we also need to analyze the
non-vacuum state expectation values for these operators. However, we
can obtain that for the non-vacuum state expectation values of these
operators, similar as that of Sec. 4 for the operator $\varphi({\bf
x},t)\star\varphi({\bf x},t)$, their results are the same as the
corresponding vacuum state expectation values. The conclusions of
microcausality for the non-vacuum state expectation values of these
operators are also the same as that for the vacuum state expectation
values of these operators. Therefore we need not to write down them
explicitly.

  To summarize, we have obtained the conclusion that the
microcausality properties for the quadratic operators $\pi({\bf
x},t)\star\pi({\bf x},t)$, $\partial_{i}\varphi({\bf
x},t)\star\partial_{i}\varphi({\bf x},t)$, and
$\partial_{i}\varphi({\bf x},t)\star\pi({\bf x},t)$ are the same as
that for the quadratic operator $\varphi({\bf x},t)\star\varphi({\bf
x},t)$. For the case $\theta^{0i}=0$ of spacetime noncommutativity,
they satisfy the microcausality. For the case $\theta^{0i}\neq0$ of
spacetime noncommutativity, they violate the microcausality.

\section{Conclusion}

\indent

  For the microcausality problem of quantum fields on noncommutative
spacetime, it need to be studied from several different aspects.
Because NCFTs cannot maintain particle Lorentz transformation
invariance [7], no matter whether $\theta^{\mu\nu}$ is a tensor or
not, it is necessary to investigate the properties of quantum fields
on noncommutative spacetime with respect to a subgroup of the usual
Lorentz group, which is the group $SO(1,1)\times SO(2)$ that leaves
$\theta^{\mu\nu}$ invariant [9]. Thus in this case, microcausality
of quantum fields on noncommutative spacetime is formulated with
respect to a two dimensional light-wedge [9]. In fact in this
representation, the traditional meaning of microcausality is
violated for quantum fields on noncommutative spacetime. Inside the
two dimensional $SO(1,1)$ light wedge, waves have infinite
propagation speed.

  On the other hand, to suppose that $\theta^{\mu\nu}$ is a Lorentz
tensor, NCFTs will maintain the observer Lorentz transformation
invariance [7]. Thus it is also necessary to study the
microcausality properties of quantum fields on noncommutative
spacetime with respect to their usual Lorentz invariant spectral
measures. For such a case, some of the results were obtained in
Refs. [13,14]. In Ref. [14], Greenberg have obtained the result that
for scalar field on noncommutative spacetime, microcausality is
violated generally, no matter whether $\theta^{0i}\neq 0$ or
$\theta^{0i}=0$. In this paper we have studied this problem further.
We obtain the result different from that of Ref. [14]. We obtain the
result that for free scalar field on noncommutative spacetime,
microcausality is satisfied when $\theta^{0i}=0$, and violated when
$\theta^{0i}\neq0$. The difference between the results of Ref. [14]
and this paper lies in several reasons, such as the form of the
Fourier expansion of scalar field and the criterion of
microcausality violation. In Ref. [14], scalar field is expanded
with respect to positive frequency (annihilation) part only. We
consider that it is not the complete Fourier expansion for quantum
fields. On the other hand, in Ref. [14], some of the results are
based on the commutators of two different operators. However as
pointed out in Sec. 2 of this paper, the criterion of microcausality
violation should be given by the commutators of the same operator,
while not two different operators. These reasons result the
different conclusions.

  Besides, we need to investigate microcausality properties of
interacting fields on noncommutative spacetime. Recently in Ref.
[15], from the generalized Bogoliubov-Shirkov criterion of causality
violation, Haque and Joglekar have obtained that for the Yukawa
interaction in noncommutative spacetime, causality is violated for
both $\theta^{0i}\neq 0$ and $\theta^{0i}=0$. In addition we can see
from Refs. [16-19] that quantum and classical nonlinear
perturbations have infinite propagation speed in noncommutative
spacetime. While these phenomena should also have relations with the
violation of causality of quantum fields on noncommutative
spacetime.

  We need also to mention here that because what we have studied in
this paper are the microcausality properties of free fields, we have
only calculated the correlation functions, i.e., the expectation
values of the Moyal commutators, while not analyzed their relations
with the $S$-matrix amplitudes. On the other hand, because we
consider that in noncommutative spacetime, the basic product
operation is the Moyal star-product, we need to investigate the
commutators of the Moyal star-products. Although what we have
studied in this paper are the microcausality properties of free
fields, the results are not trivial and obvious.

  The problem of microcausality violation of quantum fields on
noncommutative spacetime is very important, because it is related
with the existence of infinite propagation speed of physical
information as analyzed in Sec. 2 of this paper. Thus the violation
of microcausality for quantum fields on noncommutative spacetime may
have important applications in future information transmission.

  As demonstrated in Refs. [29,30], unitarity of the
$S$-matrix is lost for NCFTs with $\theta^{0i}\neq0$. However, for
NCFTs even if unitarity is satisfied, they still have many other
propertied different from field theories on commutative spacetime.
Thus for the unitarity violation of NCFTs with $\theta^{0i}\neq0$,
it may need to be explained and understood from different
approaches. Thus one may not exclude the case of time-space
noncommutativity from their unitarity problems. In fact, it is more
reasonable that time and space are in the equal position. They
should be both quantized under a very small microscopic scale. On
the other hand, for the unitarity problem of NCFTs with
$\theta^{0i}\neq0$, some authors have argued that they can be
resolved through many different methods [31-33].

\vskip 2.5 cm

\noindent {\large {\bf References}}

\vskip 20pt

[1] H.S. Snyder, Phys. Rev. {\bf 71}, 38 (1947).

[2] A. Connes, {\sl Noncommutative geometry} (Academic Press,
    New York, 1994).

[3] S. Doplicher, K. Fredenhagen, and J.E. Roberts, Phys. Lett. B
    {\bf 331}, 39 (1994);

    ~~~ Commun. Math. Phys. {\bf 172}, 187 (1995), hep-th/0303037.

[4] N. Seiberg and E. Witten, J. High Energy Phys. 09 (1999) 032,
hep-th/9908142.

[5] M.R. Douglas and N.A. Nekrasov, Rev. Mod. Phys. {\bf 73}, 977 (2001),
hep-th/0106048.

[6] R.J. Szabo, Phys. Rep. {\bf 378}, 207 (2003), hep-th/0109162.

[7] S.M. Carroll, J.A. Harvey, V.A. Kostelecky, C.D. Lane, and
    T. Okamoto, Phys. Rev.

    ~~~ Lett. {\bf 87}, 141601 (2001), hep-th/0105082.

[8] M. Chaichian, P.P. Kulish, K. Nishijima, and A. Tureanu, Phys.
    Lett. B {\bf 604}, 98

    ~~~ (2004), hep-th/0408069.

[9] L. {\' A}lvarez-Gaum{\' e} and M.A. V{\' a}zquez-Mozo, Nucl.
    Phys. {\bf B668}, 293 (2003),

    ~~~ hep-th/0305093.

[10] L. Alvarez-Gaum{\' e}, J.L.F. Barb{\' o}n, and R. Zwicky,
     J. High Energy Phys. 05 (2001)

     ~~~~~ 057, hep-th/0103069.

[11] D.H.T. Franco and C.M.M. Polito, J. Math. Phys. {\bf 46},
     083503 (2005),

     ~~~~~ hep-th/0403028.

[12] C.S. Chu, K. Furuta, and T. Inami, Int. J. Mod. Phys. A {\bf
     21}, 67 (2006),

     ~~~~~ hep-th/0502012.

[13] M. Chaichian, K. Nishijima, and A. Tureanu, Phys. Lett. B {\bf
        568}, 146 (2003),

     ~~~~~ hep-th/0209008.

[14] O.W. Greenberg, Phys. Rev. D {\bf 73}, 045014 (2006),
     hep-th/0508057.

[15] A. Haque and S.D. Joglekar, hep-th/0701171.

[16] S. Minwalla, M. Van Raamsdonk, and N. Seiberg, J. High Energy
     Phys. 02 (2000)

     ~~~~~ 020, hep-th/9912072.

[17] M. Van Raamsdonk, J. High Energy Phys. 11 (2001) 006,
     hep-th/0110093.

[18] A. Hashimoto and N. Itzhaki, Phys. Rev. D {\bf 63}, 126004
     (2001), hep-th/0012093.

[19] B. Durhuus and T. Jonsson, J. High Energy Phys. 10 (2004) 050,
     hep-th/0408190.

[20] M. Chaichian, K. Nishijima, and A. Tureanu, Phys. Lett. B {\bf
     633}, 129 (2006),

     ~~~~~ hep-th/0511094.

[21] M. Chaichian, P. Pre{\u s}najder, and A. Tureanu, Phys. Rev.
     Lett. {\bf 94}, 151602 (2006),

     ~~~~~ hep-th/0409096.

[22] Z.Z. Ma, hep-th/0601046, hep-th/0601094.

[23] A. Gerhold, J. Grimstrup, H. Grosse, L. Popp, M. Schweda, and
     R. Wulkenhaar,

     ~~~~~ hep-th/0012112.

[24] A. Micu and M.M. Sheikh-Jabbari, J. High Energy Phys. 01 (2001)
     025,

     ~~~~~ hep-th/0008057.

[25] C. Itzykson and J.-B. Zuber, {\sl Quantum field theory}
     (McGraw-Hill Inc., 1980).

[26] J.D. Bjorken and S.D. Drell, {\sl Relativistic quantum fields}
     (McGraw-Hill, 1965).

[27] L.D. Landau and E.M. Lifshitz, {\sl Quantum mechanics}
     (Pergamon Press, 1977).

[28] G. de Barra, {\sl Measure theory and integration} (Halsled
     Press, New York, 1981).

[29] J. Gomis and T. Mehen, Nucl. Phys. {\bf B591}, 265 (2000),
hep-th/0005129.

[30] A. Bassetto, F. Vian, L. Griguolo, and G. Nardelli, J. High
     Energy Phys. 07 (2001)

     ~~~~~ 008, hep-th/0105257.

[31] D. Bahns, S. Doplicher, K. Fredenhagen, and G. Piacitelli,
     Phys. Lett. B {\bf 533}, 178

     ~~~~~ (2002), hep-th/0201222.

[32] C.S. Chu, J. Lukierski, and W.J. Zakrzewski, Nucl. Phys. {\bf
     B632}, 219 (2002),

     ~~~~~ hep-th/0201144.

[33] N. Caporaso and S. Pasquetti, J. High Energy Phys. 04 (2006)
     016, hep-th/0511127.

\end{document}